\documentclass[12pt]{article}
\usepackage{amsfonts}

\usepackage{amsfonts}
\usepackage{amssymb}
\usepackage{amscd}

\def\D{\hbox{D\kern-.73em\raise.25ex\hbox{-}\raise-.25ex\hbox{ }}}
 \def\d{\hbox{d\kern-.33em\raise.75ex\hbox{-}\raise-.75ex\hbox{}}}

\def\GGG{\frak G }
\def\gr3{\GGG\,(\SSS_3)}

\def\gr2{\GGG\,(\SSS_2)}

\def\SSS{\frak S}

\def\vp{\vspace}
\def\hp{\hspace}
\def\ed{\end{document}}
\def\beq{\begin{equation}}
\def\eeq{\end{equation}}
\def\bea{\begin{eqnarray}}
\def\eea{\end{eqnarray}}
\def\ba{\begin{array}}
\def\ea{\end{array}}
\def\bi{\begin{itemize}}
\def\ei{\end{itemize}}

\def\nn{\nonumber}

\newcommand{\bp}{\noindent\begin{minipage}[c]}
\newcommand{\ep}{\end{minipage}}

\date{}

\begin{document}

 \title{\large{ ON DECOHERENCE  IN NONCOMMUTATIVE PLANE WITH PERPENDICULAR
 MAGNETIC FIELD  }}

\author{{\large{Branko
Dragovich$^1$\,\footnote{\,dragovich@phy.bg.ac.yu}} \,\, and
Miroljub Dugi\'c$^2$ }  \\ {} \\
 {\it $^1$Institute of Physics, P.O. Box 57,   11001
Belgrade, } \\{\it Serbia and Montenegro} \\ {\it $^2$Department
of Physics, Faculty of Sciences,}\\ {\it Kragujevac, Serbia and
Montenegro }}

\maketitle

\begin{abstract}
In the last years noncommutative quantum mechanics has been
investigated intensively. We consider the influence of magnetic
field  on decoherence of a system in the noncommutative quantum
plane. Particularly, we point out a model in which the magnetic
field allows {\it in situ} dynamical control of decoherence as
well as, in principle, observation of  noncommutativity.

\end{abstract}

\bigskip

\bigskip


\bigskip

{\bf PACS nukbers}: 11.10.Nx, 03.65.Yz

\bigskip

\bigskip

{\bf 1. Introduction}

\bigskip

\noindent Thinking about a possible solution of the problem of
ultraviolet divergences, already in the 1930s Heisenberg
conjectured that position coordinates might be noncommutative
(NC). Snyder \cite{snyder} was the first who started to develop
this idea systematically in 1947. An intensive interest to NC
quantum theories emerged after observation of noncommutativity in
string theory with D-branes in 1998. Most of the research has been
done in the framework of NC field theory (for reviews, see e.g.
\cite{nekrasov} and \cite{szabo}). NC quantum mechanics (NCQM) has
been also actively investigated. It enables construction of simple
NC models which have relevance to concrete phenomenological
systems and  can be regarded as the corresponding one-particle
nonrelativistic sector of NC quantum field theory. Although
spacetime uncertainties  have their origin in string theory and
quantum gravity at the Planck scale, they may play a significant
role in some important quantum-mechanical phenomena. An
experimental observation of noncommutativity at the
non-relativistic level, would be a striking physical event. To
this end, many quantum-mechanical effects  (Aharonov-Bohm effect,
lowest Landau level, fractional quantum Hall effect, ...) in a NC
background have been studied and the corresponding influences were
calculated (see, e.g. \cite{nekrasov}, \cite{szabo},
\cite{dragovich1} and references therein). Quantum decoherence is
one of the new phenomena, of fundamental and practical importance,
which combined with noncommutativity might become of the larger
interest.

Here, we investigate the possible influence of noncommutativity on
the occurrence of the  so-called decoherence effect [5-7]. Quantum
decoherence is sometimes considered to be a fundamental physical
basis for the "transition from quantum to classical" [5-10], i.e.
 the (semi)classical behaviour of the (open) quantum systems
(that cannot in principle be described by the Schr\"odinger
equation). Actually, the environment induces the effective {\it
superselection rules} for an open system by destroying the linear
("coherent") superpositions of certain states (the so-called
"pointer basis" states) of the open system [6]. The
non-interfering states thus appear "objectively" to be present for
an observer, very much like one would expect for the macroscopic
(classical) systems.

Quantum decoherence is a distinguished theory of modern quantum
mechanics. It is therefore {\it per se} interesting to investigate
the occurrence of decoherence in NCQM. At the first sight, it is
unlike to be able to observe a classical-like effect which  has
its origin in the Planck scale quantum physics. However, this
should not be the case of NCQM with decoherence. As a result, we
find some interesting consequences of noncommutativity for a
simple model of the decoherence theory. Particularly, we point out
a situation in which the experimenter may {\it in situ} (at will)
control the decoherence, i.e. by external magnetic field induce
wanted behaviour of the open system, in principle allowing the
observation of the phase space noncommutativity. A proposal for
the experimental observation of such an effect is a matter of the
work in progress. Up to our best knowledge, an influence of the
noncommutativity on decoherence has not been explored so far.

\bigskip

{\bf 2. A phase space noncommutativity}

\bigskip

\noindent We use here NCQM  which is based on the following
algebra for the position and the momentum coordinates
\cite{dragovich1}: \bea\hp{-4mm} [ \hat{x_a},\hat{p_b}] =
i\,\hbar\, (\delta_{ab} - \theta_{ac} \sigma_{cb}/4), \quad
[\hat{x_a},\hat{x_b}] = i\,\hbar\, \theta_{ab},\quad
[\hat{p_a},\hat{p_b}] = i \, \hbar\, \sigma_{ab}, \label{1} \eea
where $(\theta_{ab}) = {\Theta}$ and $(\sigma_{ab}) ={\Sigma}$ are
the antisymmetric matrices with constant elements. It allows
simple reduction to the usual algebra \bea\hp{-5mm} [
\hat{q_a},\hat{k_b}] = i\,\hbar\, \delta_{ab}, \qquad\quad
[\hat{q_a},\hat{q_b}] = 0,\qquad\quad [\hat{k_a},\hat{k_b}] = 0 ,
\label{2} \eea using the following linear transformations: \bea
\hat{ x_a} = \hat{q_a} - \frac{\theta_{ab}\, \hat{k_b}}{2}\,,
\qquad\qquad
 \hat{ p_a} = \hat{k_a} + \frac{\sigma_{ab}\, \hat{q_b}}{2}\, , \label{3} \eea
where summation over repeated indices (here and at other places)
is understood.
  In the sequel we
often take $\,\theta_{ab}=\theta\, \varepsilon_{ab}\,$ and
$\,\sigma_{ab}=\sigma\, \varepsilon_{ab},\,$ where \bea \hp{-5mm}
\varepsilon_{ab} = \left\{ \begin{array}{lll}
\hspace{2.5mm}1\,, \quad a<b \\
\hspace{2.5mm}0\,, \quad a=b \\
-1\,, \quad a>b \, .\end{array} \right\}  \label{4} \eea  \vp{2mm}

\noindent and we assume that both $\theta$ and $\sigma$  take the
very small values.

\bigskip

{\bf 3. On decoherence}

\bigskip

\noindent Quantum decoherence is a fundamental quantum effect
referring to the open systems ($S$) that are in unavoidable
interaction with their environments ($E$). The composite system
$S+E$ is described by the Hamiltonian:

\begin{equation}
\hat H  = \hat H_S + \hat H_E + \hat H_{int}
\end{equation}

\noindent where $\hat H_S$ and $\hat H_E$ represent the
self-Hamiltonians of the subsystems $S$ and $E$, while $\hat
H_{int}$ represents the interaction in the system. The composite
system is assumed to evolve in time according to the following
equation for the density matrix:
\begin{equation}
\hat \rho_{SE}(t) = \hat U(t, t_{\circ}) \hat \rho_{SE}(t_{\circ})
\hat U^{\dag}(t, t_{\circ})
\end{equation}

\noindent where $\hat U$ represents the unitary operator of (the
Schr\"odinger) evolution in time of the isolated quantum system $S
+ E$.

Now, the open system state is defined by  "tracing out" the
environmental degrees of freedom:
\begin{equation}
\hat \rho_{S}(t) = tr_E \hat \rho_{SE}(t)
\end{equation}

The decoherence (or the "{\it environment-induced superselection
rules}" [6]) effect is defined by the following two conditions
[8-10]: (i) disappearance of the off-diagonal terms of $\hat
\rho_{S}$ in a certain orthonormalized basis, the so-called
"pointer basis", $\{\vert m\rangle_S\}$:
\begin{equation}
lim_{t \to \infty}{}_S\langle m\vert \hat \rho_S (t) \vert
m'\rangle_S= 0 , \quad m \neq m'
\end{equation}

\noindent where the limit $t \to \infty$ should not be literally
understood,  and (ii) robustness of the "pointer basis" states:

\begin{equation}
\hat H_{int} \vert m\rangle_S \vert \chi \rangle_E = \vert
m\rangle_S \vert \phi \rangle_E , \quad \forall{m}
\end{equation}

\noindent for arbitrary initial state of the environment, $\vert
\chi \rangle_E $. Alternatively, decoherence is defined by the
very existence of the so-called "pointer observable" that
represents the center of algebra of the observables of the system
$S$:

\begin{equation}
\hat \Lambda_S =  \lambda_n \hat P_{Sn}.
\end{equation}

\noindent The projectors $\hat P_{Sn}$ define an orthogonal
decomposition of the Hilbert state space of the system, $H_S$:

\begin{equation}
H_S = \oplus_n H_n
\end{equation}

\noindent such that the subspaces $H_n$ represent the {\it
supereselection sectors}: due to the interaction with the
environment, the linear superpositions of the pointer basis states
belonging to the different superselection sectors are (usually
quickly) destroyed--decohered. This is the reason for referring to
$\hat \Lambda_S$ as to a "macroscopic observable", which brings
the ({\it semi}){\it classical behaviour} of the open system $S$
[5, 6, 10]; the typical macroscopic observables are the
center-of-mass coordinates. Physically, the macroscopic
observables determine the system behaviour very much like the
classical variables determine behaviour of the macroscopic bodies
[5, 10].

Investigating the occurrence of decoherence, one should take into
account the total Hamiltonian  (5). Nevertheless, the effect of
decoherence {\it never takes place} if the interaction term $\hat
H_{int}$ is not of a rather special kind [8, 9]. E.g., the
condition (ii) implies diagonalizability of $\hat H_{int}$ in the
pointer basis,
\bea\hp{-4mm} _S\langle m \vert \hat H_{int} \vert
m' \rangle_S = 0 \eea
while $\vert m \rangle_S \in H_n$ and $\vert
m' \rangle_S \in H_{n'}$, $n \neq n'$. The approximate equality in
(12) gives rise to the approximate pointer basis, i.e. to the
approximate equalities in (8) and (9).

The expression  (12) is simultaneously a definition and the
condition of existence of the pointer basis, thus defining the
superselection sectors $H_n$. Therefore, existence of the pointer
basis is a necessary, but not  a sufficient condition for the
occurrence of decoherence. In the sequel, we refer to the task of
investigating the existence of the pointer basis.

Decoherence takes some time that is characterized by the so-called
decoherence time, $\tau_D$, which is typically of the form, e.g.
(cf. (13) below) $\tau_D \sim g_{ij}^{-1}$, where $g_{ij}$
represent the coupling constants of the interaction in  composite
system. By definition, exceedingly long time intervals point out
non-occurrence of decoherence, while, typically, $\tau_D$ is a
very short interval (e.g.  a particle of the mass  $1$g, under the
standard macroscopic conditions and for the spatial distances of
the order of $1$cm,  decoheres in $10^{-23}$s [10]).

\bigskip

{\bf 4. The model}

\bigskip

\noindent We assume the usual linear interaction of the open
system with its environment \bea \hat{H}_{int} = g_{ij}\,
\hat{x}_i \, \hat{C}_j + f_{pq}\, \hat{p}_p \, \hat{D}_q,
\label{5} \eea where $\hat x_i$ and $\hat p_i$ are the position
and momentum coordinates of a quantum system, for the virtually
arbitrary observables of the environment. For simplicity, we omit
the subscripts $S$ and $E$ for the system and environment, while
one assumes the tensor product of the observables: e.g. $\hat x \,
\hat C \equiv \hat x \otimes \hat C$, where $\hat x$  and $\hat C$
refer to the system and environment, respectively.

Let a charged particle with its charge $e$ be moving in a NC plane
with commutation relations (\ref{1}). Using transformations
(\ref{3}), one obtains \bea  \hat{H}_{int} = g_{ij}\, (\hat{q}_i -
\frac{\theta}{2} \, \varepsilon_{il} \, \hat{k}_l ) \, \hat{C}_j +
f_{ij}\, (\hat{k}_i  + \frac{\sigma}{2} \, \varepsilon_{il} \,
\hat{q}_l ) \, \hat{D}_j , \label{6} \eea where $\hat{q}_i$ and
$\hat{k}_i$ satisfy commutation relations (\ref{2}).

Interaction with magnetic field $\mathcal{B}$ can be introduced by
replacement $\hat{k}_l \to \hat{k}_l - e \, \hat {\mathcal A}_l ,$
where electromagnetic potential $\hat \mathcal A_l =
\frac{\mathcal B}{2} \, \varepsilon_{lm}\, \hat{q}_m$.  Then we
have \bea  \nn \hat{H}_{int} =  g_{ij}\, \Big(\hat{q}_i  -
\frac{\theta}{2} \, \varepsilon_{il} \, (\hat{k}_l - \frac{e\,
{\mathcal B}}{2} \, \varepsilon_{lm} \, \hat{q}_m ) \Big) \,
\hat{C}_j \\ + f_{ij}\, \Big(\hat{k}_i  - \frac{e \,\mathcal B}{2}
\, \varepsilon_{im} \, \hat{q}_m + \frac{e\, {\sigma}}{2} \,
\varepsilon_{im} \, \hat{q}_m  \Big) \, \hat{D}_j , \label{7} \eea
where magnetic field is perpendicular to the NC plane. Since in
two-dimensional case $\varepsilon_{il}\, \varepsilon_{lm} = -
\delta_{im}$, we obtain \bea \nn \hat{H}_{int} = \hat q_i \Big( (1
- e \,\mathcal B \, \theta /4)\, g_{ij} \, \hat C_j - (e \,
\mathcal B /2 - \sigma/2)\,  f_{pq} \, \varepsilon_{pi} \, \hat D_q \Big)\\
+ \hat k_i \, \Big( f_{ij} \, \hat D_j - (\theta /2)\,
\varepsilon_{pi} \, g_{pq} \, \hat C_q \Big)  \,. \label{8} \eea

Existence of the magnetic field allows, in principle, the
possibility of the external control of decoherence in the system
modeled by  (16). This is the subject of the next section.

\bigskip

{\bf 5. The pointer basis}

\bigskip

\noindent Here, we refer solely to investigating diagonalizability
(12) of $\hat H_{int} $, i.e. existence of the pointer basis, that
is a necessary condition for the occurrence of decoherence. In
general, the occurrence of decoherence requires the analysis of
the total Hamiltonian (5).

\bigskip

{\bf 5.1 The "macroscopic considerations"}

\bigskip

\noindent In his classic textbook, von Neumann [11] introduced the
"coarse graining" of the position- and the momentum- axes, thus
allowing the approximations of these observables. In fact, the
observable $\hat q_i$ is approximated by the corresponding
observable $\hat \xi_i$, while the momentum coordinate $\hat k_j$
is approximated by $\hat \pi_j$, so that one has \bea \Vert (\hat
q_i - \hat \xi_i) \vert \Psi_{\mu\nu}\rangle\Vert \le 60 \, \Delta
\hat x_i \eea and analogously for $\hat k_j$. The point is that,
given $\Delta \hat q_i \, \Delta \hat k_j = \delta_{ij} \, \hbar
/2$, the observables $\hat \xi_i$ and $\hat \pi_j$ may satisfy:
(a) $[\hat \xi_i, \hat \pi_j] = 0, \forall{i, j}$, and (b) the
common eigenbasis $\{\vert \Psi_{\mu\nu}\}$ can be obtained from
the orthonormalization of the minimal-uncertainty states, with the
pure discrete spectra for both, $\hat \xi_i$ and $\hat \pi_j$.

Now, the model  (16) can be rewritten in the form \bea \hat
H_{int} = \hat \xi_i \, \hat E_i  + \hat \pi_i \, \hat F_i + \hat
H' \eea where \bea \hat E_i = (1 - e \mathcal B \theta/4)\, g_{ij}
\, \hat C_j - (e \mathcal B /2 - \sigma/2) \, f_{pq} \,
\varepsilon_{pi} \, \hat D_q \eea and \bea \hat F_i = f_{ij} \,
\hat D_j - (\theta /2)\, \varepsilon_{pi} \, g_{pq} \, \hat C_q
\eea while, due to (17), one may write \bea \Vert \hat H'\Vert \ll
\Vert (\hat H_{int} - \hat H') \Vert. \eea

Therefore, due to (21), the basis $\{\vert \Psi_{\mu\nu}\}
\rangle$ appears as the approximate pointer basis yet for the
price of large standard deviation of both, $\Delta \hat q_i \sim
10 \sqrt{\hbar}$, and $\Delta \hat k_j \sim 10 \sqrt{\hbar}$.

Needless to say, these "large" uncertainties may become useful in
the macroscopic context of the theory, i.e. as regards the large,
macroscopic bodies. However, if the open system ($S$) with
characteristic dimension $L$ is such that $L \ll \Delta \hat q_i
$, the system position becomes completely undetermined
(un-decohered) -- as we learn from the experiments on the
fullerene spatial interference [12].

The decoherence time for the model  (18) is obviously of the
order of $\{min[\vert f_{ij}\vert, \vert g_{ij}\vert]\}^{-1}$.

\bigskip

{\bf 5.2 The coordinate coupling}

\bigskip

\noindent Let us assume that \bea f_{pq} = 0,\quad \forall{\, p,
q}\eea thus redefining the model  (13) to \bea \hat H_{int} =
g_{ij} \hat x_i \hat C_j. \eea Without noncommutativity ($\theta =
0 = \sigma$) the position eigenbasis $\vert \vec q \rangle ,$
where $\vert \vec q \rangle = \vert \vec x \rangle$ if $\theta =
0$, appears as the exact pointer basis for the system. Then,
obviously, $\tau_D \sim \{min[\vert g_{ij}\vert]\}^{-1}$.

However,  taking the model  (16) into account, the condition (22)
gives rise to the following redefinition of the Hamiltonian, \bea
\hat H'_{int} = (1 - e\mathcal B \theta /4)\, g_{ij} \,\hat q_i \,
\hat C_j - (\theta /2)\, \varepsilon_{pi} \, g_{pq} \, \hat k_i \,
\hat C_q \approx (1 - e\mathcal B \theta /4)\,  g_{ij} \, \hat q_i
\hat C_j ,\eea since $\theta$ is assumed to take a very small
value.

Then, the basis $\vert \vec q \rangle$ appears as the approximate
pointer basis, for the decoherence time  $\tau_D \sim \vert 1 - e
\,\mathcal B \, \theta /4\vert^{-1} \,\{min[\vert
g_{ij}\vert]\}^{-1}$.

\bigskip

{\bf 5.3 The momentum coupling}

\bigskip

\noindent Let us now assume that \bea g_{pq} = 0 , \quad
\forall{\, p, q}\eea thus redefining the model  (13) to \bea \hat
H_{int} = f_{ij} \, \hat p_i \, \hat D_j . \eea In the absence of
noncommutativity  for $\mathcal{B} = 0$, the momentum eigenbasis
$\vert \vec k \rangle ,$  where $\vert \vec k \rangle = \vert \vec
p \rangle$ if $\sigma = 0$,   appears to be the exact pointer
basis for the system. Then, $\tau_D \sim \{min[\vert
f_{ij}\vert]\}^{-1}$.

However,  taking the model  (16) into account, the condition
 (25) gives rise to the following redefinition of the
Hamiltonian, \bea \hat H_{int} = - (e\mathcal B /2 - \sigma /2)\,
f_{pq}\, \varepsilon_{pi} \, \hat q_i \, \hat D_q + f_{ij} \, \hat
k_i \, \hat D_j. \eea This is an interesting case, indeed.

Actually, if $\mathcal B = 0$ or $\mathcal B \ll 1$, the model
 (27) reads \bea \hat H'_{int} \approx f_{ij} \, \hat k_i \, \hat
D_j \eea thus giving rise to $\vert \vec k \rangle$ as the
approximate pointer basis, cf. (26), and $\tau_D \sim \{min[\vert
f_{ij}\vert]\}^{-1}$.

However, for strong magnetic field, $\mathcal B \gg 1$, one
obtains \bea \hat H''_{int} \approx - (e\mathcal B /2 )\, f_{pq}
\, \varepsilon_{pi} \, \hat q_i \, \hat D_q  \eea thus giving rise
to the approximate pointer basis $\vert \vec q \rangle$, while the
decoherence time $\tau_D \sim \vert e\, \mathcal{B}\vert^{-1} \,
\{\min[\vert f_{ij}\vert]\}^{-1}$. The form (29) equally refers to
both (16) and (26) for the sufficiently strong magnetic field

Therefore, choosing the appropriate  external magnetic field, one
may obtain the mutually {\it exclusive} physical situations: for
weak magnetic field, the open system reveals its (approximate)
momentum $\vert\vec k \rangle$, while for sufficiently strong
magnetic field, the system reveals its (approximate) position,
$\vert\vec q \rangle$.

\bigskip

{\bf 5.4 Comments}

\bigskip

\noindent Even as the approximate pointer states, the states
$\vert \vec q \rangle$ or $\vert \vec k \rangle$ give rise to
better resolution of the system position or momentum than dealing
with the "macroscopic states" $\vert \Psi_{\mu\nu} \rangle$. In
principle, the states $\vert \vec q \rangle$ (or $\vert \vec k
\rangle$) refer to the {\it exact} position (or momentum) of the
system, up to the Planck scale. Of course, the details in this
subject depend on both, the details in the interaction as well as
in the self-Hamiltonian of the system $S+E$.

Interestingly enough, as long as the linear coupling is of
interest, the possible occurrence of decoherence allows in
principle partial distinguishing between the models discussed in
the previous subsections. To this end, the situation is as
follows:

\smallskip

(a) Appearance of the pointer states $\vert \vec q \rangle$ in the
presence of a strong magnetic field gives rise to either the
general model (13), or  the model (29). Then, the decoherence
time is proportional to $ \mathcal{B}^{-1}$.

\smallskip

(b) Appearance of the pointer states $\vert \vec k \rangle$ for
the weak field, points out the relevance of the model (28).

\smallskip

(c) Appearance of the pointer states $\vert \vec q \rangle$
independently on the magnetic field points out the relevance of
the model (24).

\smallskip

(d) Non-appearance of either of the pointer states $\vert \vec q
\rangle$ or $\vert \vec k \rangle$ (independently on the magnetic
field) points out the general model (13), i.e. (18).

\bigskip

{\bf 5.5 Revealing noncommutativity}

\bigskip

\noindent The conclusions of the sections 5.1 through 5.4 equally
refer to both, the "commutative" models ($\theta = 0 = \sigma$) as
well as to the "noncommutative" models ($\theta \neq 0 \neq
\sigma$). This is due to the very small values of $\theta$ and
$\sigma$. The only instant, generally, where the two cases can be
mutually distinguished, i.e. where the noncommutativity may reveal
itself through the possible occurrence of the decoherence effect,
is determined by the following condition \bea \mathcal{B} =  4/e
\theta \eea \ in which case the model (16) reads (compare to
(29)) \bea \hat H_{int} \approx  (e\mathcal B /2 )\, f_{pq} \,
\varepsilon_{pi}\,  \hat q_i \, \hat D_q  = - (2 / e\theta) \,
f_{pq} \, \varepsilon_{pi} \, \hat q_i \, \hat D_q. \eea

In this case, the position eigenstates $\vert \vec q \rangle$
appear as the approximate pointer basis states, for the
decoherence time $\tau_D $ proportional to $  e \,\theta$. Bearing
in mind a small value of $\theta$, we conclude that this situation
requires a very strong (probably physically unrealistic) magnetic
field yet giving rise eventually to  fast decoherence.

\bigskip

{\bf 6. Physical analysis of the model}

\bigskip

\noindent Of special interest is the following physical
situation. An open system experiences decoherence due to the
interaction of the type (13). Now, the external magnetic field is
applied.

By manipulating the external magnetic field, one can in principle
perform {\it in situ} change of the system behaviour. E.g., cf.
Section 5.3, initially, the system reveals its momentum yet for
the price of the totally undetermined position; in such a
situation, one may observe the spatial interference for the
system in analogy with the spatial interference of the fullerene
molecules [12]. Now, by applying the sufficiently strong magnetic
field, one may (in principle) obtain the "appearance" of the
system in front of the experimenter's eyes (yet for the price of
the loss of information about the particle momentum). The effect
may be visualized in analogy with the production of the
(spatially) localized electronic states in atoms [13]. Needless
to say, this effect would give rise to the loss of the spatial
interference, i.e. of the loss of the contrast between the
interference fringes--in analogy with the observation of the
spatial decoherence for the fullerene molecules [14].
Interestingly enough, in case of (30), this effect simultaneously
reveals the underlying phase space noncommutativity (cf. Section
5.5).

The magnetic field appears as an externally controllable
parameter. As distinct from the parameter $T$ (the temperature) in
the fullerene experiments [12, 14], the magnetic field allows
{\it in situ} control  of decoherence: the experimenter may, {\it
at will}, {\it dynamically} change the decoherence-induced
behaviour of the open system. Observing such an effect
experimentally would be very interesting, indeed; a work in this
direction is in progress. To this end, the following requirements
should be met: (i) a well-defined open system (modeled by (13))
that is confined to a plane, (ii) the possibility of making a
choice of the magnetic field, providing that the strong field
does not alter either the definition of the environment or  the
interaction in the composite system, and (iii) well-defined
procedures for determining the pointer states as well as the
decoherence time for the open system.

\bigskip

{\bf 7. Discussion and Conclusion}

\bigskip

\noindent The linear coupling (13) is both mathematically the
simplest possible as well as the best known (investigated) in the
decoherence theory so far. To this end, the two remarks are in
order. First, the model (13) extends the usual linear position
coupling, which is analysed in some detail in the literature [5,
10, 15, 16]; our model (23) reproduces the main results presented
therein. The inclusion of the momentum coupling ($f_{ij} \neq 0$)
in (13) is the ultimate basis of the observations and results
presented in this paper. E.g., we point out appearance of the
magnetic field as a parameter allowing {\it in situ} dynamical
control of the mutually {\it exclusive} (complementary), the
decoherence-induced behaviour of the open system. While this is
{\it per se} interesting--the experimenter may, at will, cause
"disappearance" and "appearance" of the open system--this kind of
the open system control essentially resembles the main idea of the
"error avoiding" methods in quantum computation theory [17-19].
Actually, driving $S$ e.g. into the momentum space ($\vert \vec k
\rangle $ are the pointer basis states) {\it keeps the coherence}
of the position states ($\vert \vec q \rangle $), and {\it vice
versa}. Second, the model (13) is the most general one as long as
there is not any linear dependence [9] in the set of the
environmental observables, $\{\hat C_i, \hat D_j\}$. We assume
the environment ($E$) is a "macroscopic" system [6-10], for which
the effects coming from the Planck scale may be  ignored. This is
the reason we do not take into account the phase space
noncommutativity for the observables of the environment, $\hat
C_i$ and $\hat D_j$.

Noncommutativity may influence the effect of decoherence {\it
only} when the external control (here: the magnetic field) is
constrained by the noncommutativity parameters $\theta$ and/or
$\sigma$. In case that  (30) is fulfilled, the following physical
effect takes place: instead of revealing only approximate,
"macroscopic", position {\it and} momentum, the particle reveals
its position $\vert \vec q \rangle$ in a very short time interval
of the order of $\theta$. In all other cases--Section 5.1 through
5.4--noncommutativity is hidden  due to the small values of the
parameters $\theta$ and $\sigma$. E.g., in the case of  "momentum
coupling" (Section 5.3), we obtain the possibility to observe  a
striking physical effect: by manipulating the external magnetic
field, the open system may reveal {\it either} its position {\it
or} its momentum. The possible occurrence of decoherence may, in
principle, partly reveal the type of interaction as long as the
linear model (13) is in question. This possibility is stressed by
the points (a)-(d) in Section 5.4.

Therefore, in the context of the model (13), by manipulating the
external magnetic field, one may obtain: (a) a partial information
about the type of interaction in the composite system, and/or (b)
the possibility of the external, {\it in situ} (dynamical),
control of decoherence, and/or (c) the open system decoherence
reveals the phase space noncommutativity. As to the point (c), the
magnetic field $\mathcal{B}$ resembles the noncommutativity--cf.
the term $\mathcal{B}/2 - \sigma /2$ in (16). It is therefore not
for surprise that, for $\mathcal{B} = 0$ or $\mathcal{B} \ll 1$,
the cases for which $\theta \neq 0 \neq \sigma$ appear
essentially indistinguishable from the cases for which $\theta =
0 = \sigma$.

We conclude that, for a system that is an object of decoherence,
the phase space noncommutativity may influence the occurrence of
decoherence, in the sense of the possible redefinition (or even
change) of the pointer basis and/or of the decoherence time. To
this end, certain external control of decoherence is required, cf.
(30). Our analysis of a comparatively simple model
 (13) encourages investigation towards the more realistic
physical models.

\bigskip

\noindent {\bf Acknowledgements\,} The work on this paper was
supported  by Ministry of Science and Environmental Protection,
Serbia, under contract No 1426.

\end{document}